\def\kms{km~s$^{-1}$}
\def\lsun{L$_{\odot}$}

\def\msun{M$_{\odot}$}

\def\aap{A\&A}

\def\apj{ApJ}
\def\aj{AJ}

\def\mnras{MNRAS}

\documentclass{iaus}
\usepackage{graphicx}

\title[High resolution observations of BA supergiants] %% give here short title %%
{High spatial resolution monitoring of the activity of BA supergiant winds}

\author[O. Chesneau et al.]   %% give here short author list %%
{Olivier Chesneau$^1$
 \and L. Dessart$^2$
 \and A. Kaufer$^3$
 \and D. Mourard$^1$
 \and O. Stahl$^4$
 \and R. Prinja$^5$
 \and S. Owocki$^6$
 }

\affiliation{$^1$UMR 6525 H. Fizeau, Univ. Nice Sophia Antipolis, CNRS, Observatoire de  la C\^{o}te d'Azur, Av. Copernic, F-06130 Grasse, France \\ email: {\tt olivier.chesneau@oca.eu} \\[\affilskip]
$^2$Laboratoire d'Astrophysique de Marseille, Universit\'e de Provence,
CNRS, 38 rue Fr\'ed\'eric Joliot-Curie, F-13388 Marseille Cedex 13,  
France \\[\affilskip]

$^3$European Southern Observatory, Alonso de Cordova 3107, Casilla 19001, Santiago 19, Chile\\[\affilskip]
$^4$Zentrum f\"ur Astronomie der Universit\"at Heidelberg, Landessternwarte,
K\"onigstuhl 12, 69117 Heidelberg, Germany\\[\affilskip]
$^5$Department of Physics \& Astronomy, University College London, Gower Street, London, WC1E 6BT, UK\\[\affilskip]
$^6$Bartol Research Institute, Dept. of Physics \& Astronomy, Univ. of Delaware Newark, DE 19716 USA

}

\pubyear{2008}
\volume{xxx}  %% insert here IAU Symposium No.
\pagerange{119--126}
\jname{Title of your IAU Symposium}
\editors{A.C. Editor, B.D. Editor \& C.E. Editor, eds.}
\begin{document}

\maketitle

\begin{abstract}
There are currently two optical interferometry recombiners that can provide spectral resolutions better than 10000, AMBER/VLTI operating in the H-K bands, and VEGA/CHARA, recently commissioned, operating in the visible. These instruments are well suited to study the wind activity of the brightest AB supergiants in our vicinity, in lines such as H$\alpha$ or Br$\gamma$. We present here the first observations of this kind, performed on Rigel (B8Ia) and Deneb (A2Ia). Rigel was monitored by AMBER in two campaigns, in 2006-2007 and 2009-2010, and observed in 2009 by VEGA; whereas Deneb was monitored in 2008-2009 by VEGA. The extension of the H$\alpha$ and Br$\gamma$ line forming regions were accurately measured and compared with CMFGEN models of both stars. Moreover, clear signs of activity were observed in the differential visibility and phases. These pioneer observations are still limited, but show the path for a better understanding of the spatial structure and temporal evolution of localized ejections using optical interferometry.
\end{abstract}

\firstsection % if your document starts with a section,
\section{Introduction}

Supergiants of spectral types B and A (BA-type supergiants) are evolved massive stars of typical initial mass of 25-40\,\msun\ and high luminosity ($\gtrsim$10$^5$\,\lsun). Their luminosity and temperature place them among the visually brightest massive stars, a particularly interesting aspect
for extragalactic astronomy. Nearby BA supergiants have been analyzed with sophisticated radiative-transfer tools, and among them the closest ones Deneb ($\alpha$\,Cygni, HD~197345, A2\,Ia) and Rigel ($\beta$\,Orionis, 
HD~34085, B8Ia).

Most BA supergiants rotate slowly ($v \sin i$ of about 25-40km\,s$^{-1}$), at least relative to their terminal wind-velocity of $\sim$200-400\,\kms (\cite[Fraser et al. 2010]{2010MNRAS.404.1306F}, \cite[Lefever et 
al. 2007]{2007A&A...463.1093L}). Intensive spectroscopic monitoring of the activity of some wind sensitive lines, such as H$\alpha$ has lead to the conclusion that variability in the stellar winds of luminous hot stars is localized and structured and rotates around the stars. Obviously, the winds are modulated, w by patches
  on the stellar surfaces produced either by non-radial pulsation
  (NRP) patterns or magnetic surface structures. 

\section{VEGA/CHARA observations in the visible}
The VEGA recombiner of the CHARA array (Mt Wilsin, CA) is a recently commissioned facility that provides spectrally dispersed interferometric observables, with a spectral resolution reaching R =30 000, and a spatial resolution of less than one $mas$ \cite[Mourard et 
al. (2009)]{2009A&A...508.1073M}. The instrument recombines currently the light from two telescopes, but 3-4 telescope recombination modes are foreseen in a near future. As an example, the extension of the H$\alpha$ line-forming region of the prototypical Herbig star AB\,Aur was resolved for the first time \cite[Rousselet-Perraut et 
al. (2010)]{2010A&A...516L...1R}.

The H$\alpha$ line of bright, slow rotators such as Deneb or Rigel can be isolated from the continuum, and the spatial properties of the line-forming region can thus be studied with unprecedented resolution. Using the smallest baseline of the CHARA array (baseline of 34m), we conducted a pioneering temporal monitoring of Deneb uncovering a high level of activity in the H$\alpha$ line-forming region. Rigel was also observed a few times. 
The information in a line was extracted differentially by comparing the properties of the fringe between a reference 
channel centered on the continuum of the source, and a sliding science narrow channel.

For the most accurate estimates to date of the diameters of Rigel and Deneb, we refer to \cite[Aufdenberg et al. (2008)]{2008poii.conf...71A} in which CHARA/FLUOR observations in the K band with baselines reaching 300\,m are described. 
These observations infer a UD angular diameter of 2.76$\pm$0.01\,mas, and 2.363$\pm$0.002\,mas, 
for Rigel and Deneb, respectively.

\section{New perspectives for estimating the mass-loss rate}

Many useful diagnostics of mass loss are used from the radio to the X-rays, based on the determination of the  free-free continuum emission at radio
wavelengths, the fitting of some line emission (in particular H$\alpha$,  ultraviolet (UV)
resonance-line absorption or or more recently the
fitting of X-ray spectra (Cohen et al. these proceedings). These estimators suffer from different biases depending on whether their physical mechanisms is linearly or quadratically dependent on the local density. They also suffer different degrees of contamination and require different ancillary knowledge of, e.g., the excitation or ionization conditions in the wind, the velocity
structure of the wind, or the distance to the star.

We discuss here some new ways to constrain the consistency of the mass-loss rate or the wind velocity by measuring accurately the limb-darkening of massive close-by BA supergiants photosphere, and the extension of the line forming regions of some wind-sensitive lines such as H$\alpha$. Of course, these estimations are not free from their own biases, although they are based from observables that are very different from those that are commonly used.

Radiative-transfer calculations were carried out with the line-blanketed non-LTE 
model-atmosphere code {\sc CMFGEN}. In this study, we used the stellar parameters obtained by \cite[Przybilla et 
al. (2006)]{2006A&A...445.1099P} and \cite[Schiller 
\& Przybilla (2008)]{2008A&A...479..849S}  for  Rigel and Deneb, respectively. 
An important conclusion is that the second lobe of the visibility curve is far more sensitive to any fluctuation of the mass-loss rate in the visible than the infrared. The reasons are twofold: a higher sensitivity to limb-darkening effects in the visible, and a more extended continuum-formation region, despite the very limited amount of flux involved (the wind remains in any case optically thin). Balmer bound-free cross-sections increase from 400 to 800nm. This causes the continuum photosphere to shift weakly in radius across this wavelength range and also alters the limb-darkening properties of the star. 

\begin{figure}
% \vspace*{-2.0 cm}
\begin{center}
 \includegraphics[width=11cm]{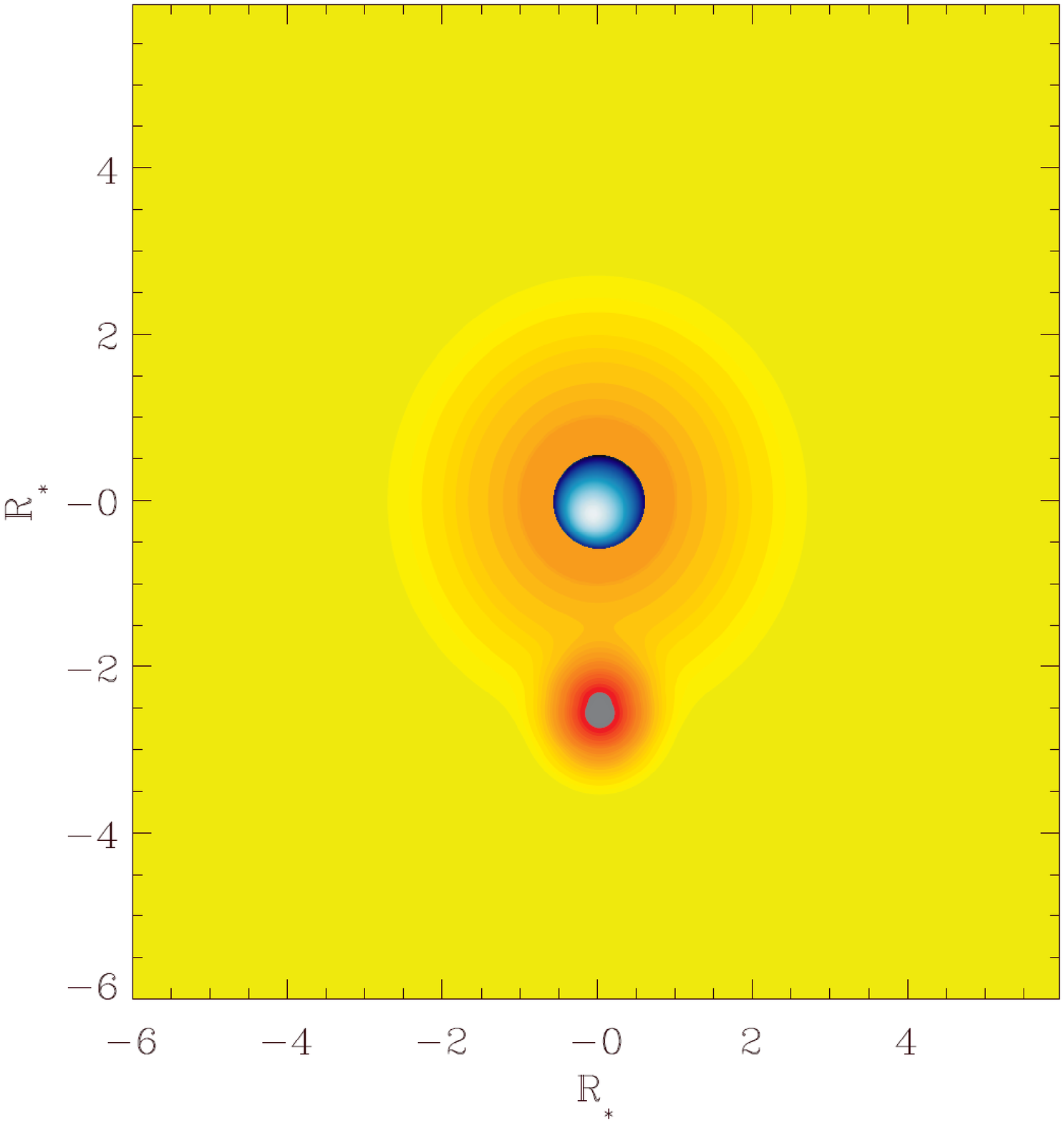} 
% \vspace*{-1.0 cm}
 \caption{Illustrative picture summarizing the observations from FLUOR/CHARA \cite[(Aufdenberg et al. 2006)]{2006AAS...208.0601A}, providing the low inclination fast-rotator photospheric model and the VEGA/CHARA observation of July 2008, showing a the H$\alpha$ emission around the star. The picture was not consistently generated. The H$\alpha$ envelope was created in the frame of a non-rotating CMFGEN, and the perturbation  artificially created by locally multiplying the flux by a coefficient following a Gaussian distribution. The FWHM of the Gaussian is about half the radius of the star, and the flux increase at peak is 10 times the local flux. The position, FWHM and flux contrast of the perturbation were chosen to fit the interferometric data. }
   \label{fig1}
\end{center}
\end{figure}

\section{AMBER/VLTI observations in the near-IR}
An interferometric monitoring campaign of Rigel has been undertaken at the ESO/Paranal observatory with the
the near-infrared instrument of the VLTI using the highest spectra resolution (R=12000) and baselines reaching 125m. Two periods of monitoring were performed in 2006-2007 and in 2009-2010. These observations were completed by intensive spectroscopic observations using the FEROS and BESE spectrographs\footnote{This is a FEROS copy running at the 1.5m Bochum telescope on Armazones}. The data are under analysis and only preliminary results are discussed here.
The first test was to apply the model of Rigel used to account for the two observations of Rigel in H$\alpha$ with VEGA/CHARA. The Br$\gamma$ visibilities are much deeper than predicted by the model, and the mass-loss rate must be significantly increased to match the observations. This can hardly be explain by the intrinsic variability of the stars and the use of non-contemporary observations, and might be an indication that the model is not consistently accounting for the visible AND near-IR observations.

With VEGA/CHARA, a significant S-shape differential phase signal was observed once with the North-South baseline. This S-shape signal can be interpreted as a rotation of the H$\alpha$ line forming region.
The latest dynamical spectrum of H$\alpha$ from the BESO data set shows that a large scale event is seen in the H$\alpha$ around JD2455240 (12-02-2010). A clear change in the profile of the Br$\gamma$ line as seen by AMBER is visible, corresponding to a marked photocenter offset in the red part of the line indicating that the ejected material was located in a particular direction at a defined radial velocity.

Both the rotation and the kinematics of the ejection event are not currently accounted for by the CMFGEN models developed so far, that do not include any kinematics apart from the wind velocity law. This is not an issue in the case of Deneb which is probably seen at large inclination, and it was possible to account for the large phase signal affecting the full H$\alpha$ line by a perturbation of the model. However, it is not possible with such a non-rotating model to generate a perturbation that would affect only part of the line. This is why a refined rotating model for Rigel is necessary.

\section{On the fast rotation of BA supergiants}
Aufdenberg et al. presented evidence that Deneb is a fast rotator, based on 24 high accuracy 
FLUOR/CHARA visibilities in the $K'$-band with projected baselines ranging from 106\,m to 310\,m, sampling the first and second lobes 
of the visibility curve. They detected slight departures from a purely spherical model at a level of about 2\% in the near-IR, a discovery 
that had not been anticipated for an AB supergiant. They tentatively fitted the data with a rotating model atmosphere, and demonstrated that despite 
the low $v\sin i$ of the star, a model at half critical speed, seen nearly pole-on may account for the 
interferometric observations. That no evidence of rotation was detected in the differential phases is an additional argument for a low $v\sin i$, and therefore a nearly pole-on configuration for Deneb. Such a signal was observed on Rigel by VEGA in H$\alpha$ and by AMBER in Br$\gamma$ which suggests that Rigel is seen at significantly higher inclination, with an axis direction approximately oriented north-south. Note that the $vsini$ of Deneb and Rigel is about 20\kms and 35\kms, respectively.

Rapidly rotating evolved supergiants appear to be relatively rare and one might wonder whether they might share some relationship with B[e] star. As discussed in Fraser et al. 2010, there are at least three channels by which these stars could have reached their current evolutionary state: (1) they had extremely high rotational velocities on the main sequence, and have spun down normally as they evolved; (2) they have not spun down as much as most other stars as they became supergiant stars; (3) they had a moderate velocity on the main sequence but have experienced an additional source of angular momentum that has allowed them to maintain their rotational velocity, and this source might be a close companion.

The example of the A[e] supergiant HD\,62623 is very informative in this context (see Millour et al., Meilland et al. these proceedings). This is the coolest member of the so-called B[e] stars, surrounded by dense, dusty disks. HD62623 exhibits a rich emission line spectrum and only few photospheric lines are observable, alike many B[e] stars. Spectroscopically and spatially resolved observations of AMBER/VLTI have shown that the supergiant lies in a cavity, and is surrounded by a dense disk of plasma. The Br$\gamma$ line from the supergiant is {\it in absorption} showing that the star is not much different from a normal member of its class such as Deneb, albeit with a significantly large (but not overwhelming) $vsini$ of about 50\kms. By contrast, the Balmer and Bracket lines are much larger ($vsini$ of about 120\kms), and the AMBER observations demonstrated that they originate from a disk of plasma, most probably in Keplerian rotation (Millour, Meilland et al. submitted). It is difficult to understand how HD\,62623 would exhibit a much larger mass-loss rate compared to its class, and how such a dense equatorial disk could have been generated. However, high quality spectral monitoring performed more than 15yrs ago showed that a companion star rotates close to the supergiant with a period of about 136 days \cite[Plets et 
al.(1995)]{1995A&A...293..363P}. The spectroscopic signal is weak, hence the mass ratio is very large, and the companion is probably a solar mass star (given that the recent interferometric observations have provided invaluable constraints on the inclination of the system). This tightens further the connection between B[e] stars and binarity  

The recent Spitzer observations of the mid-IR excess observed in LMC B[e] stars provide further evidence of the analogy to the class of post-asymptotic giant branch stars with binary companions and dusty, circumbinary disks. The SEDs are all similar, within the LMC group of B[e] stars and also compared with galactic onces, leading \cite[Kastner et al. (2010)]{2010AJ....139.1993K} to quote this sentence in their abstract:'we speculate that B[e] supergiant stars may be post-red supergiants in binary systems with orbiting, circumbinary disks that are derived from post-main-sequence mass loss' .

\section{Prospects}
The observations reported here are a pioneering effort in order to improve upon the spectral and spatial resolution of the interferometric facilities. An extension of the VEGA observations is possible for sources with an apparent angular diameter between 0.5 and 1.5\,mas,  that are large enough, but also bright enough to make use of the highest spectral dispersion of the VEGA instrument (limiting 
magnitude of about 3). This concerns the AB supergiants closer than 1.5 kpc, and the supergiants in the Orion complex ($d \sim$500\,pc) are in this context of particular interest. Simultaneous 3 telescope recombination would provide much better $uv$ coverage than the one presented in this paper. The number of sources accessible to AMBER observations are more restricted as the spatial resolution in the near-IR is decreased by about 4, and no baselines longer than 130m are to date accessible. Moreover, the Br$\gamma$ line is formed closer to the photosphere. Pa$\beta$ is the H band is better suited, but the best line by far would be the HeI1080 line in the J band. Long baselines FLUOR/CHARA observations of Rigel and Deneb should be repeated to confirm that these measures are independent of the star activity (\cite[Miroshnichenko 2007]{2007ApJ...667..497M}). The B[e] star with the largest infrared excess of the class, HD\,87643 appeared also, in view of the AMBER/VLTI observations, as a unique and quite extreme large separation binary system. The large nebula that surrounds this system shows periodic arc that suggest a possible link between extreme period of mass ejection and the periastron passages of the companion with a highly eccentric orbit (\cite[Millour et 
al. 2009]{2009A&A...507..317M}).

\end{document}